\documentstyle[epsf]{mn2e}

\title [Challenging times: a re-analysis of NGC 5408 X-1]{Challenging times: a re-analysis of NGC 5408 X-1}

\author[M.\,J. Middleton, T.\,P. Roberts, C. Done \& F.\,E. Jackson]
{M.\,J. Middleton, T.\,P. Roberts, C. Done \& F.\,E. Jackson\\
Department of Physics, University of Durham, South Road, Durham
DH1 3LE,
UK\\
}

\date{}
\pagerange{\pageref{firstpage}--\pageref{lastpage}} \pubyear{2010}

\begin{document}

\topmargin = -0.5cm

\maketitle

\label{firstpage}

\begin{abstract}

The ultraluminous X-ray source (ULX), NGC 5408 X-1, is one of only 3
such objects to show a quasi-periodic oscillation (QPO) in its power
spectrum. Previous analysis of this signal identified it with the
well-studied type C low-frequency QPO (LFQPO) seen in black hole
binaries (BHBs), implying an intermediate mass black hole
(IMBH). However, in BHBs this QPO has a centroid frequency which
scales tightly with the position of the low-frequency break in the
broad band power spectrum. We use this relation to predict the
frequency of the power spectral break in NGC 5408 X-1, and show that
this is inconsistent with the break frequencies in both available,
archival {\it XMM-Newton} observations. Thus the broad band power
spectral shape does not support this identification of the QPO.

The energy spectra also do not support an IMBH interpretation. They
can be fit by a two-component model, best described by soft thermal
emission at low energies, together with low-temperature, optically
thick Comptonisation producing a tail which dominates above 2~keV. The
parameters of the tail are unlike those seen in any of the
sub-Eddington BHB spectral states. The energy dependent variability
supports this deconvolution, as it is consistent with the soft thermal
component below 2~keV diluting extreme variability of the high energy
tail. The only objects with similar spectra which have similar amounts
of variability are the BHB, GRS 1915+105, and some extreme NLS1s.
This suggests that NGC 5408 X-1 is in a similar super-Eddington state,
placing a natural limit on the mass of $\leq$ 100M$_{\odot}$. Its QPO
could then be similar to the ultra-LFQPO seen occasionally in GRS
1915+105, consistent with a large stellar mass black hole.  We suggest
a model geometry which may explain the spectra and variability of
highly super-Eddington sources.

\end{abstract}

\begin{keywords} accretion: accretion discs -- black hole physics -- X-rays: binaries -- X-rays: individual: NGC 5408 X-1 

\vspace{-7mm}
\end{keywords}

\section{Introduction} 

Extragalactic X-ray sources with luminosities in excess of
10$^{39}$ erg s$^{-1}$ have now been widely detected in the nearby
Universe (Fabbiano 1989; Miller \& Colbert 2004; Roberts 2007). These
are designated as ultraluminous X-ray sources (ULXs) if they are not
coincident with the nucleus of their host galaxy (i.e. an AGN) nor
correspond to a foreground or background object, e.g. a QSO. They are
too luminous to be powered by a sub-Eddington accretion flow in a
stellar-mass black hole binary (BHB) unless substantial beaming of
their emission is invoked (King et al. 2001), so either require
super-Eddington mass accretion onto a stellar mass black hole, or
sub-Eddington accretion onto an intermediate mass black hole (IMBH;
see Colbert \& Mushotzky 1999). Unambiguously distinguishing between
these possibilities requires radial velocity measurements of the
underlying binary motion to directly determine the mass (see
e.g. Charles \& Coe 2006). Until these become available, we can only
use more indirect arguments to identify the nature of these objects,
primarily by comparing the X- ray spectral and timing properties of
ULXs with those of stellar-mass BHBs (see e.g. the review by Roberts
2007).

The X-ray spectra of ULXs can be fairly well described by a two
component (disc plus power-law) model, as commonly used for
BHBs. Taking the disc temperatures at face value implies masses of
order 10$^{2}$--10$^{5}$M$_{\odot}$ (Kaaret et al. 2003; Miller et al. 2003; Miller et
al. 2004a; Cropper et al. 2004; Dewangan et al. 2004; Roberts et
al. 2005), favouring sub-Eddington accretion flows onto an IMBH
(Colbert \& Mushotzky 1999). However, the disc component generally
does not dominate these spectra, and BHBs show that in these
circumstances the derived disc radii vary dramatically, so cannot be
used to estimate mass (Kubota \& Done 2004). Kajava \& Poutanen (2009)
show that this 'cool disc' component varies as $L\propto T^{-3.5}$,
implying that the radius of the thermally emitting material increases
with increasing luminosity, in sharp contrast to the $L\propto T^4$
expected from a disc with constant inner radius (see also Feng \& Kaaret 2009).  The observed
behaviour may instead be produced from a super-Eddington source
powering a strong wind, where the photosphere of the wind expands with
increasing luminosity (Shakura \& Sunyaev 1973; King \& Pounds 2003;
Begelman, King \& Pringle 2006; Poutanen et al. 2007).

While the 'disc' spectrum is ambiguous, the 'power-law' is somewhat
clearer in distinguishing between IMBH and stellar mass systems (or
rather in distinguishing between sub and super-Eddington
accretion). Sources with the highest quality data show clear signs
that the high energy `power-law' tail rolls over above 6~keV, so it is
better fit by low temperature thermal Comptonisation (Stobbart,
Roberts, \& Wilms 2006; Miyawaki et al. 2009). A rollover at such low
temperatures is not seen in any of the sub-Eddington BHBs, so this
suggests that ULXs are super-Eddington and
in a new, {\em ultraluminous} state (Roberts 2007; Gladstone et al. 2009, hereafter
G09; Vierdayanti et al.  2010). Similarly, low temperature thermal
Comptonisation is also required in the brightest Narrow-Line Seyfert
1s (NLS1s, e.g. RE J1034+396) and the brightest BHB in our own Galaxy, GRS
1915+105 (Zdziarski et al. 2001; Middleton et al. 2009; Ueda et
al. 2009). Super-Eddington accretion then implies a much smaller mass
black hole in ULXs, of the order 10-100~M$_{\odot}$.

Another diagnostic of the nature of the accretion flow is its rapid
variability as this is strongly correlated with spectral
state. Sub-Eddington BHBs show power density spectra (PDS) which can
be very roughly described by band limited noise, with a low frequency
break at $\nu_b$ from a slope of $\nu^{0}$ to $\nu^{-1}$, followed by
a high frequency break, $\nu_{h}$, to $\nu^{-2}$.  This broad-band
noise is often accompanied by strong low frequency Quasi-Periodic Oscillations (LFQPOs) which correlate with the low frequency
break ($\nu_{LFQPO} \approx 10 \nu_b$: Wijnands \& van der Klis
1999). The LFQPO increases in frequency, coherence and power as the
source flux increases and the spectrum softens from a faint low-hard
state through the hard-intermediate state up to the soft-intermediate
state (Belloni, M{\'e}ndez \& Homan 2005; van der Klis 2004; Remillard
\& McClintock 2006). During this transition, the broad-band power
spectrum narrows in $\nu P_\nu$, with $\nu_b$ increasing while the
power at high frequencies remains fairly constant (Czerny et al. 2008;
Done, Gierli{\'n}ski \& Kubota 2007).

\begin{figure}
\begin{center}
\begin{tabular}{l}
\leavevmode \epsfxsize=8cm \epsfbox{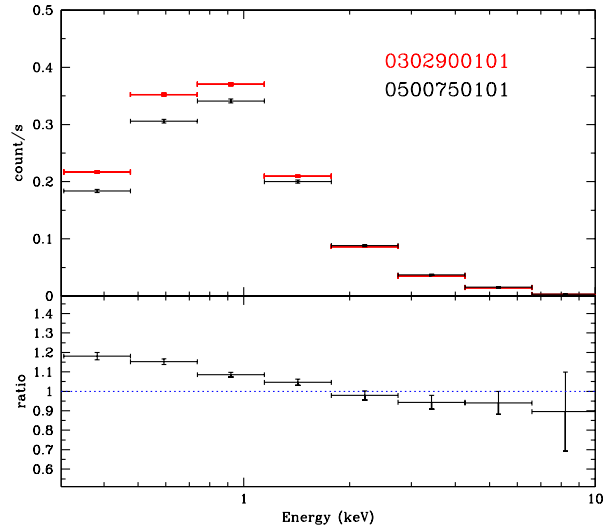}
\end{tabular}
\end{center}
\caption{Mean count rate in each observation over the course of each
respective GTI in the lightcurve. The ratio of the
observations (Obs 1/Obs 2) demonstrates that count rate significantly
drops between observations at soft energies below $\sim$1 keV but is consistent with staying
the same at energies above this.}
\label{}
\end{figure}

Whilst the characteristic breaks are difficult to constrain in the low
signal-to-noise ULX data (see Heil, Vaughan \& Roberts 2009), QPO
features are much easier to measure. Three ULXs are known to
demonstrate strong QPOs in their PDS, M82 X-1 (Strohmayer \& Mushotzky
2003), NGC 5408 X-1 (Strohmayer et al. 2007, hereafter S07; Strohmayer
\& Mushotzky 2009, hereafter SM09) and, more recently, a second source
in M82 (Feng, Rao, \& Kaaret 2010). Identifying these with the type C LFQPO
seen in sub-Eddington BHBs implies that these are all IMBHs (Casella
et al. 2008; S07; SM09). However, it also implies that the power
spectra should show a low frequency break, and that the energy spectra
should be consistent with a low-hard or very high state as these are
the states in which the LFQPO is seen. We reanalyse the data from NGC
5408 X-1, looking at both spectra and variability, and show that
neither the energy spectra nor power spectra support the identification of
the QPO with the standard LFQPO seen in sub-Eddington BHBs. Instead we
show that the spectra, power spectra, and 
variability as a function of energy are all
better matched by a super-Eddington flow, implying a stellar mass
compact object, in which case the QPO may be analogous to the milli-Hz
QPOs seen in GRS 1915+105.

\begin{figure*}
\begin{center}
\begin{tabular}{l}
\leavevmode \epsfxsize=16cm \epsfbox{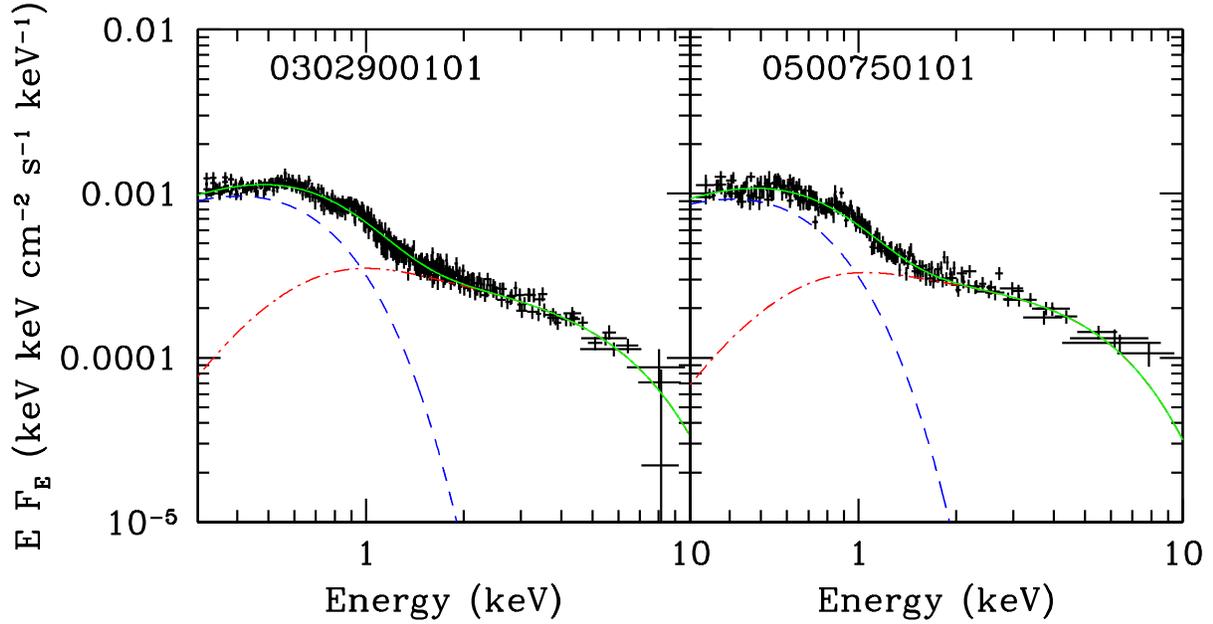}
\end{tabular}
\end{center}
\caption{Energy spectra of combined EPIC MOS and PN observations of
NGC 5408 X-1. Both spectra indicate a break around 2--3 keV or a
deficit of photons in 1--2 keV band relative to the extension of the
power-law, fitted to the data above 4 keV.  This can be interpreted as
a signature of thermal Comptonisation, thus the absorption-corrected
convolution in green (solid line) is composed of a thermal disc
component ({\sc diskpn}) in blue (dashed line), together with thermal
Comptonisation of the seed disc photons ({\sc comptt}) in red
(dot-dashed line).}
\label{}
\end{figure*}

\begin{table*}
\begin{center}
\caption{Table of best-fitting spectral parameters}
\begin{minipage}{185mm} 
%\bigskip
\begin{tabular}{l|c|c|c|c|c|c|c|c|c|c}
  \hline
  
& \multicolumn{4}{|c|} {{\sc diskpn}} & \multicolumn{4}{|c|} {{\sc comptt}} & \\
  \hline

  % after \\: \hline or \cline{col1-col2} \cline{col3-col5} ...
   & $N_{\rm H}$  & $T_{\rm max}$ (keV) & norm  & F$_d$ & $T_{\rm e}$ (keV) & $\tau$   & norm  &  F$_c$ & $\chi^{2}$ \\

 & & & ($\times10^{-3}$) & & &  & ($\times10^{-4}$) & & \\

 \hline

Obs 1 & 9.6 $_{-0.5}^{+0.8}$ & 0.17 $_{-0.01}^{+0.00}$ & 2.9 $_{-0.3}^{+0.1}$ & 1.6 $_{-0.4}^{+0.7}$ & 1.6 $_{-0.2}^{+0.4}$ & 6.3 $_{-1.1}^{+0.9}$ & 5.8 $_{-8.7}^{+7.7}$  & 1.2 $_{-0.5}^{+1.1}$ & 959 (829 d.o.f.)  \\ 

Obs 2 &  10.8$_{-1.3}^{+1.0}$  & 0.17 $_{-0.00}^{+0.01}$ & 2.7 $_{-0.9}^{+0.5}$ & 1.6 $_{-0.6}^{+1.2}$ & 1.4 $_{-0.3}^{+0.5}$ & 7.0 $_{-1.5}^{+2.2}$ & 5.9 $_{-1.2}^{+1.0}$ & 1.2 $_{-0.6}^{+2.0}$ & 637 (589 d.o.f.) \\

\hline

\end{tabular}

\end{minipage}
\end{center}
\begin{minipage}
\textwidth{} {Notes: Best-fit parameters for the model: {\sc
tbabs*(diskpn+comptt)}. $N_{\rm H}$ ($\times$10$^{20}$cm$^{-2}$) is
the neutral absorption column with the lower limit set at Galactic
(5.73$\times$10$^{20}$cm$^{-2}$), $T_{\rm max}$ is the peak disc
temperature (with $R_{\rm in}$ set at 6$R_{\rm g}$) which provides the
seed photons for the electron plasma at temperature $T_{\rm e}$,
$\tau$ is the optical depth of the same plasma and F is the 0.3-10 keV
flux ($\times$10$^{-12}$ erg cm$^{-2}$ s$^{-1}$) in the disc (F$_d$)
and Compton component (F$_c$) respectively. All errors are 90\%
confidence limits.}
\end{minipage}
\end{table*}

\section{Data extraction}

The source was well-exposed on 2 occasions for all 3 EPIC cameras
on-board \textit{XMM-Newton} (OBSIDs: 0302900101, hereafter Obs 1 \&
0500750101, hereafter Obs 2). These are separated by almost exactly 2
years and have a duration of 132 and 116 ks respectively.

We follow standard extraction procedure, re-processing the raw data
files in {\sc sas v9.0} with up-to-date cross-calibration files. Taking extraction
regions of 38'' for both source and background (ensuring that the
background region was within the central chip, far enough removed from
the source to avoid PSF wings and not co-incident with other sources
in the field of view), we extract lightcurves and spectra using {\sc
xselect v5.10} and using
patterns 0-4 and 0-12 for PN and MOS respectively. The extracted lightcurves are found to be
contaminated by background flares at the beginning and end of Obs 1,
so excluding these gives a single Good Time Interval (GTI) of
$\sim$80ks. However, for Obs 2, the background flares are more
extensive, giving only $\sim$30ks of continous, clean data.  This is in keeping
with the data used for the recent rms-flux analysis of Heil \& Vaughan
(2010) but is markedly different to that of SM09 who used $\sim$70ks
from the lightcurve in Obs 2.

Figure 1 shows the EPIC PN count rate as a function of energy for each
observation, together with its ratio (Obs1/Obs2). Despite using
different GTIs to S07 and SM09, we find a similar result, which is that
there is a 20 per cent decrease at soft energies in Obs 2 compared to
Obs 1, but that there is no significant change at hard energies.

\section{Energy spectra}

For each observation we fit the EPIC PN and MOS spectra simultaneously
in {\sc xspec 11.3.2}, allowing the normalisation of the model to
float between the different instruments. The best fitting values of
these constants are consistent with each other to within 10 per cent
in all the fits.

The spectrum of the first observation has been extensively studied by
G09, where a break was found in the power-law component at hard
energies, consistent with other high quality ULX data. They showed that
Obs 1 was in fact significantly better fitted by a disc plus low
temperature Comptonisation model than by a disc plus power-law
model. Here, we revisit and improve that analysis, and apply a similar
analysis to Obs 2 for the first time.  We follow G09 and fit both
observations with a {\sc diskpn + comptt} model, where {\sc diskpn}
has $R_{\rm in}$ fixed at 6$R_{\rm g}$.  This is absorbed by a neutral
column ({\sc tbabs}) with lower limit set to that of Galactic
absorption at 5.73$\times$10$^{20}$cm$^{-2}$ (SM09). The resulting
best-fitting models are plotted with the de-absorbed data in Figure 2, and
the associated parameters are given in Table 1 together with their
90\% confidence limits.  

The coronal component is cool ($T_{\rm e} \sim 1.5$ keV) and optically
thick ($\tau \sim 7$) in both fits, providing a physical basis for the
turn-over evident at $\sim 4$ keV in both spectra in Figure 2.  We
assess the significance of this high energy break in the following,
simple manner.  We took the best-fitting model and fixed the
temperature of the corona at (i) 100 keV, similar to the corona seen
in the hard state (Ibragimov et al. 2005); and (ii) 20 keV, similar to
that in the very high/steep power-law state (Kubota \& Done 2004).  We
then re-fitted the model.  In each case this led to a degradation in
the quality of the fit, with $\Delta\chi^2 > 28$ for one additional
degree of freedom (d.o.f.) for cases (i) and (ii) in Obs 1, and
$\Delta\chi^2 > 14$ for Obs 2, compared to the free fit resulting in
the cool corona.  Statistically, this leads to $> 99.9\%$ confidence
in all cases (according to the F-test) that the cool, optically thick
solution provides a superior fit to the hotter corona cases.  This argues
strongly that the energy spectrum cannot be well-described by classic
sub-Eddington accretion states, as assumed by SM09.

Caballero-Garcia \& Fabian (2010) highlight that the spectrum of NGC
5408 X-1 contains line-like residuals at low energies (particularly
below 1 keV; see their Figure 1).  Introducing an optically thin
plasma ({\sc mekal}) component to each best-fitting model, to account
for these features, significantly improved the fit ($\Delta \chi^2$ of 83 for 2 d.o.f.
in Obs 1 and 50 for 2 d.o.f. in Obs 2, for a solar abundance {\sc
mekal} and a best fitting temperature $\sim 0.8$ keV).  Previous work (SM09)
has suggested that an extended X-ray component may be present,
associated with the star formation regions close to NGC 5408 X-1 in
its host galaxy.  One might na{\"i}vely assume that this plasma could
originate in such a star-forming region.  However, we note the large
$L_{\rm X}$ from this plasma ($\sim$2.6$\times$10$^{38}$ erg s$^{-1}$)
may be somewhat high for a small star-forming region in a small
galaxy; for context, the whole of M101 has a diffuse $L_{\rm X}$ of
$\sim$ 2$\times$10$^{39}$erg s$^{-1}$ (Warwick et
al. 2007). Furthermore, we find minimal evidence for extended emission
in an analysis of the longest available \textit{Chandra} dataset
(OBSID:4558). This suggests that the features are intrinsic to the
source (as suggested by Caballero-Garcia \& Fabian 2010), which we
will explore further in a future paper. In this paper, we are
concerned only with the continua, and the parameters of this do not differ
significantly from the fits reported in Table 1 after the addition of
the {\sc mekal} component.

\section{Timing analysis}

\subsection{Broad band power spectra}

We extract evenly-sampled PDS for each observation using the {\sc
Xronos} tool {\sc powspec}, from the background subtracted, co-added
lightcurves of the EPIC PN and MOS detectors (using the same
continuous GTI and extraction regions as used in the energy spectral
analysis). Co-adding the data maximises the total count rate and so
maximises the power spectral statistics. While the spectral response
differs between PN and MOS detectors, the fractional variability at
any specific energy should be perfectly correlated across both
(although see Barnard et al. 2007 for possible,
extraction-method-dependent problems). 

As the observations are taken in full-frame mode, the maximum time
resolution available across all 3 detectors is that for the MOS
cameras of 2.6~s. However, in order to have sufficient statistics for
a broad-band variability analysis we use a time binning of 10~s. This
binning gives an upper limit to the frequency range (0.05 Hz), just
below the frequency at which the Poisson (white noise) errors become
dominant. 

\begin{figure}
\begin{center}
\begin{tabular}{l}
\leavevmode \epsfxsize=8cm \epsfbox{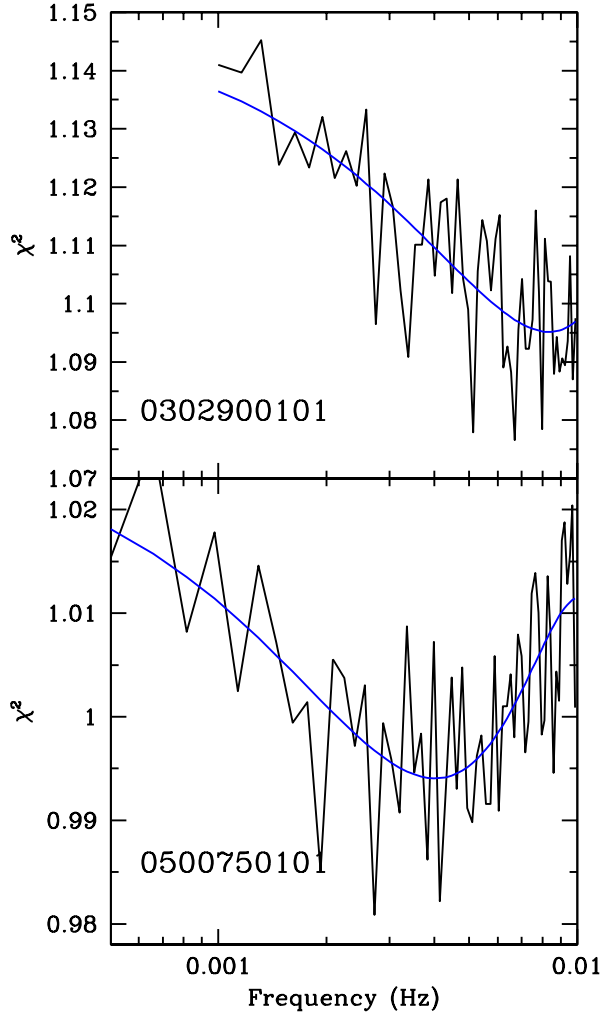}
\end{tabular}
\end{center}
\caption{Panels showing the reduced $\chi^2$ from simulating a large
number of lightcurves with the same statistical properties as the
original lightcurve. This allows us to test for the presence of a
break by including the feature in the simulations across a suitable
frequency range. In each case the residual stochasticity is averaged
by fitting a quadratic (Obs 1) or cubic (Obs 2) function (blue
lines). In both cases the well-constrained relationship between break
frequency and QPO frequency (Wijnands \& van der Klis 1999) predicts a
type C LFQPO at higher frequencies than those observed.}
\label{}
\end{figure}

\begin{figure*}
\begin{center}
\begin{tabular}{l}
\leavevmode \epsfxsize=16cm \epsfbox{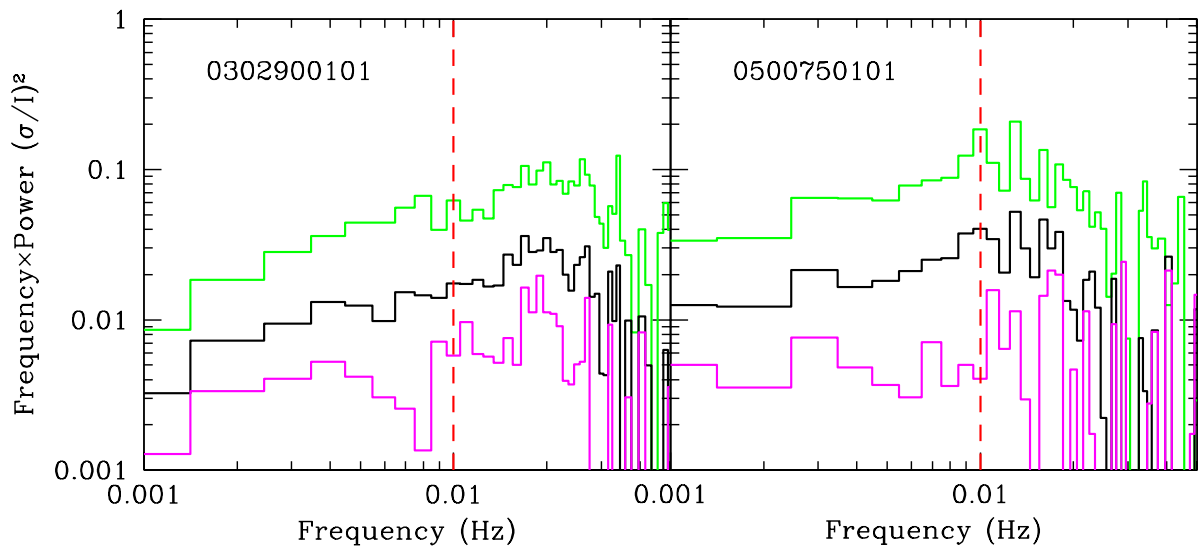}
\end{tabular}
\end{center}
\caption{The average PDS for Obs 1 (left) and 2 (right) from 1000~s
lightcurve segments in different energy bands.  These have white noise
subtracted and are geometrically re-binned by a factor of 1.02. This
only has an effect at the highest frequencies, where it reduces some
of the variance associated with white noise subtraction, but preserves
the full (linear) resolution at lower frequencies. 
Green (upper histogram in both)
corresponds to 1-8~keV, black (middle histogram in both) to 0.3-10~keV
and magenta (lower histogram in both) to 0.3-1~keV respectively. The
vertical red line indicates the upper limit on the PDS sampled by the
rms spectrum on 50s binning (Fig 5).  Plainly there is more
variability in the hard energy band than in the soft in both
datasets.}
\label{}
\end{figure*}

\begin{figure*}
\begin{center}
\begin{tabular}{l}
\leavevmode \epsfxsize=16cm \epsfbox{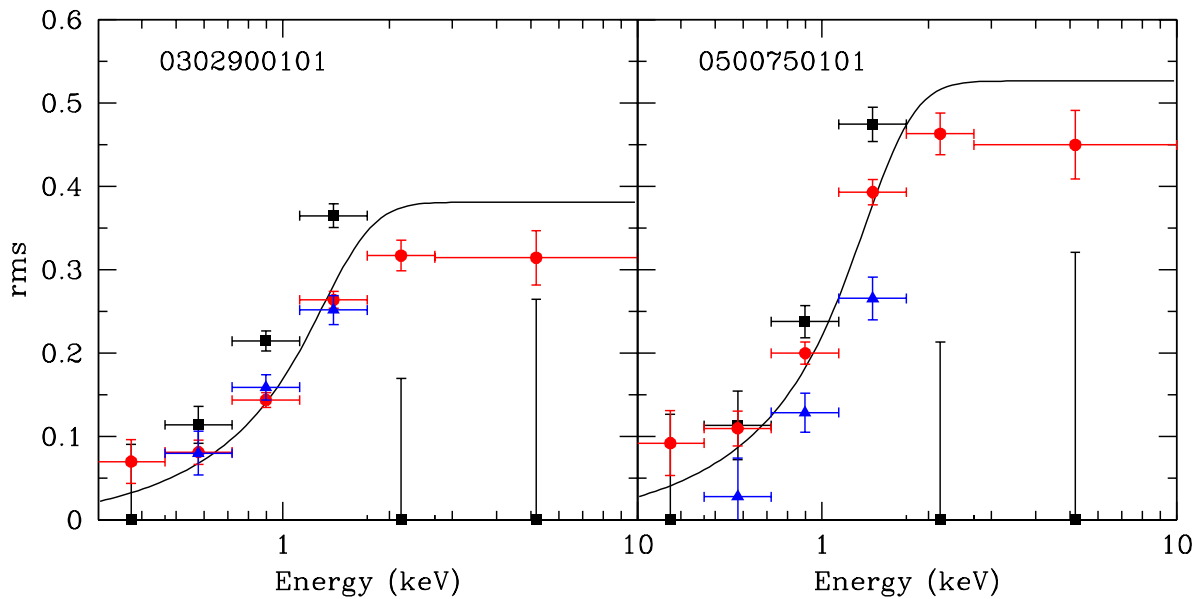}
\end{tabular}
\end{center}
\caption{The rms spectra calculated from the lightcurves in each
energy band. The black points (squares) correspond to the full frequency
bandwidth i.e. from $1/T_{obs}$ to the Nyquist frequency for 10~s
binning. It is clear that the variability rises strongly with energy
for both observations up to $\sim 1-2$~keV. After this the
signal-to-noise is very poor so the variability may either level off
or decrease with energy.  There are only marginal differences in the
pattern of variability with energy on different timescales. The red
points (circles) show the rms which results from calculating the rms from
$1/T_{obs}$ using 50~s binning of the lightcurve, so that it is more
sensitive to the longer term variability (i.e. the PDS to the left of
the vertical red line in Fig 4), while the blue points (triangles) show the
difference between these two, so that this gives the rms of the rapid
variability, integrated over frequencies from $1/(2\times 50)$ to
$1/(2\times 10)$~Hz. All these show that the amount of variability
increases with energy up to 1-2~keV, consistent with the model where
the variability at low energies is diluted by a constant soft
component (black line)}.
\label{}
\end{figure*}

\begin{figure*}
\begin{center}
\begin{tabular}{l}
\leavevmode \epsfxsize=8cm \epsfbox{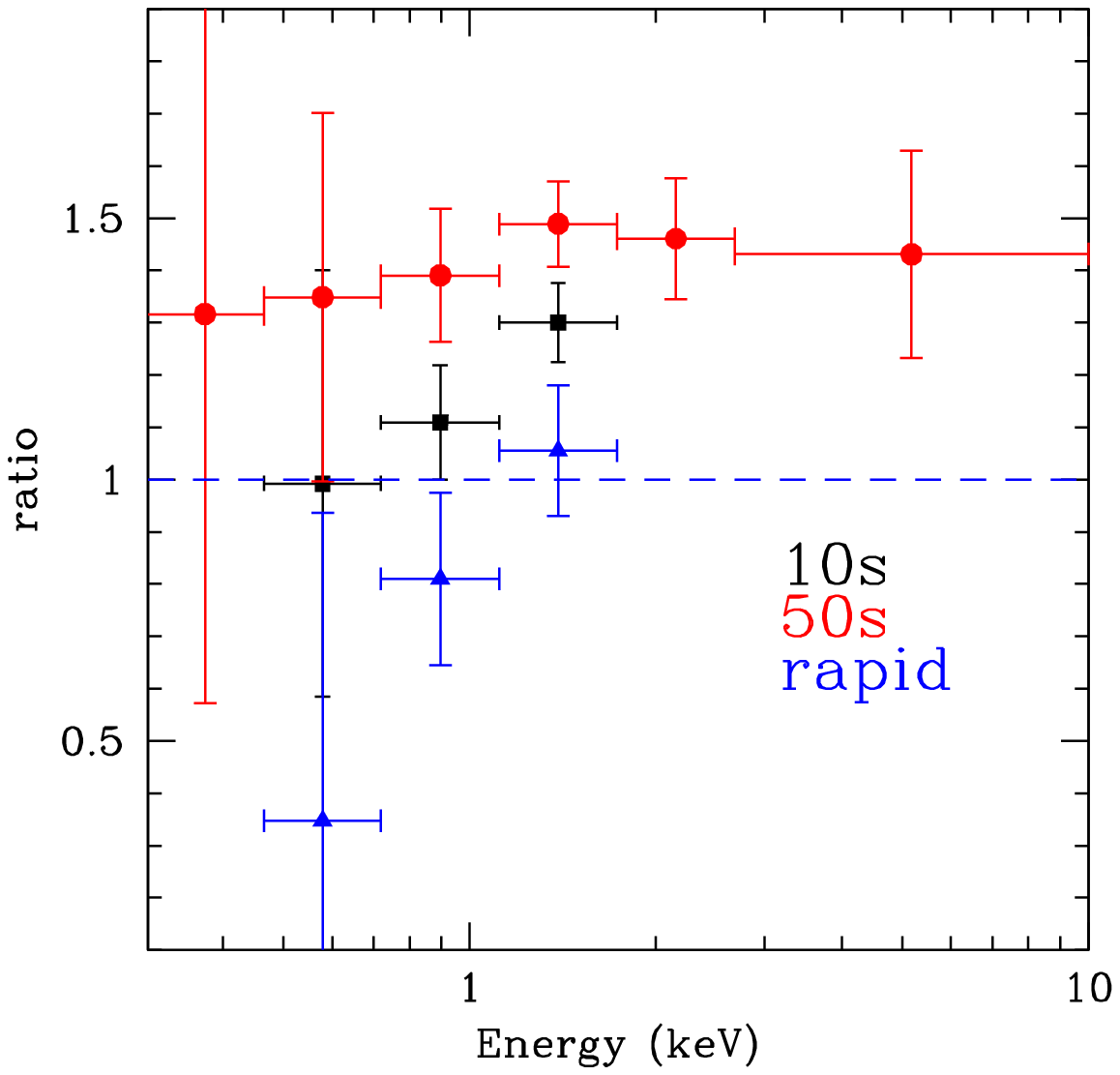}
\end{tabular}
\end{center}
\caption{Ratio of the rms between Obs 2 and Obs 1, with fractional
errors propagated through and respective colours (and shapes) corresponding to
those in Figure 5. Clearly the long timescale variability (red - circles)
increases significantly ($>$3$\sigma$) between 1-3~keV from  
Obs 1 to Obs 2.}
\label{}
\end{figure*}

S07 and SM09 show that the significance of the QPO (found at
$\sim$0.01 Hz and $\sim$0.02 Hz in Obs 1 and 2, respectively) is
maximised over the energy bands 0.2-8 keV and 1-8 keV,
respectively. We use the same energy bands to constrain the band noise
power as this generally correlates with the QPO (McClintock \&
Remillard 2006). Our power spectra show the QPO at $\sim$0.01 Hz and
$\sim$0.02 Hz for Obs 1 and 2, respectively. The significance of these
features has been confirmed in previous analyses (SM09; Heil, Vaughan
\& Roberts 2009) and so here we concentrate instead on the shape of
the broad band noise. If this QPO is the standard type C LFQPO seen in
BHBs, then the broad band noise should break from $\nu^0$ to
$\nu^{-1}$ at a frequency which is a factor $\sim$10 lower than that
of the QPO (Wijnands \& van der Klis 1999). We can test for the
presence and position of this break in our data and compare it to the
expected and observed position of the QPO. In order to do this we use
the entire length of each observation, calculating a PDS extending
down to the lowest available frequencies (1/the length of the
observation), so as to get the best possible constraints on a low
frequency break. However, power spectral statistics mean that the
uncertainty on the derived power is equal to the power itself leaving
the continuum unconstrained. We get much better constraints from
simulations, i.e. we create large numbers of simulated lightcurves
with a model power spectrum which breaks from $\nu^0$ to $\nu^{-1}$ at
a frequency $\nu_b$ which we vary from $0.001-0.01$~Hz. The simulated
lightcurves have the same statistical properties as the original,
including noise power (Middleton \& Done 2010, see also Uttley et
al. 2001), after which we obtain the average PDS of the simulated
lightcurves, together with uncertainties given by the dispersion in
the simulation values. These can then be compared to the real power
spectra to get a $\chi^2$ description of how well the simulations
describe the data, and hence constrain $\nu_b$.

The upper and lower panels in Figure 3 show reduced $\chi^2$ against
$\nu_b$ for the comparison between the simulations and the PDS for Obs
1 and 2. These are not smooth as power spectra are stochastic, but
fitting the reduced $\chi^2$ data with a quadratic allows us to
determine the 90\% significance bounds ($\Delta\chi^2=$2.7).

This gives a lower limit on the break frequency of $>6.9\times
10^{-3}$ Hz for Obs 1, while Obs 2 gives a detection at
$4.0^{+1.2}_{-1.1}\times10^{-3}$~Hz. Obs 1 is clearly inconsistent
with the predicted break at $10^{-3}$~Hz for a type C LFQPO at
0.01~Hz. The break in Obs 2 is closer to the type C prediction of $2\times
10^{-3}$~Hz, but still outside the 90 per cent confidence interval
allowed by the data. Thus the broad band power spectral shape does not
support the identification of the QPO with a type C LFQPO. 

\subsection{Energy-dependent variability}

The energy spectrum compresses all variability information into a
single average count rate, while a power spectrum compresses energy
information. Here instead we combine the two approaches, looking at
the energy dependence of the variability in order to disentangle different 
physical components in the spectrum.

Figure 4 shows the average PDS of each observation, calculated from
$\sim$80 (Obs 1) or 30 (Obs 2), 1000~s segments of the lightcurve. These
are shown in soft (0.3-1 keV, magenta), hard (1-8 keV, green) and
total (0.3-10 keV, black) band-passes, where the normalisation of the
PDS is $(\sigma/I)^2$ (where $\sigma^2$ is the intrinsic,
statistical error corrected, variance of the lightcurve and I is the
mean count rate). Plainly there is more variability at hard energies
than at soft energies, thus the total bandpass PDS lies between the
two energy-resolved PDS. There is also significantly more variability
in the hard energy band in Obs 2 than Obs 1, with $\sigma/I=0.426\pm
0.008$ for Obs 1 increasing to $0.514\pm 0.013$.

We get a more direct view of the energy dependence of the variability
by calculating the rms variability spectrum, $\sigma/I(E)$ (see
Edelson et al. 2002; Markowitz, Edelson \& Vaughan 2003; Vaughan et
al. 2003; Gierli{\'n}ski \& Zdziarski 2005). To do this, we extract the total
lightcurve over the length of the observation in each chosen energy
bandpass (0.3-0.45, 0.45-0.7, 0.7-1.1, 1.1-1.8, 1.8-2.5,
2.5-10~keV). We first use 10~s binning, so we are in effect
integrating the power spectrum in that energy band between frequencies
of $1/T_{obs}$ to the Nyquist frequency of 0.05~Hz. The results from
this are shown by the black points in Figure 5.  In each observation
there is a very significant increase in rms from soft through to
intermediate energies, peaking at an extraordinary 40\% in Obs 1 and
$\sim$50\% in Obs 2!  Following this peak, the fractional variability
appears to flatten or perhaps even decrease at higher energies, though
the lack of statistics at these energies means that this is not at all
well constrained.

\begin{figure*}
\begin{center}
\begin{tabular}{l}
\leavevmode \epsfxsize=17.5cm \epsfbox{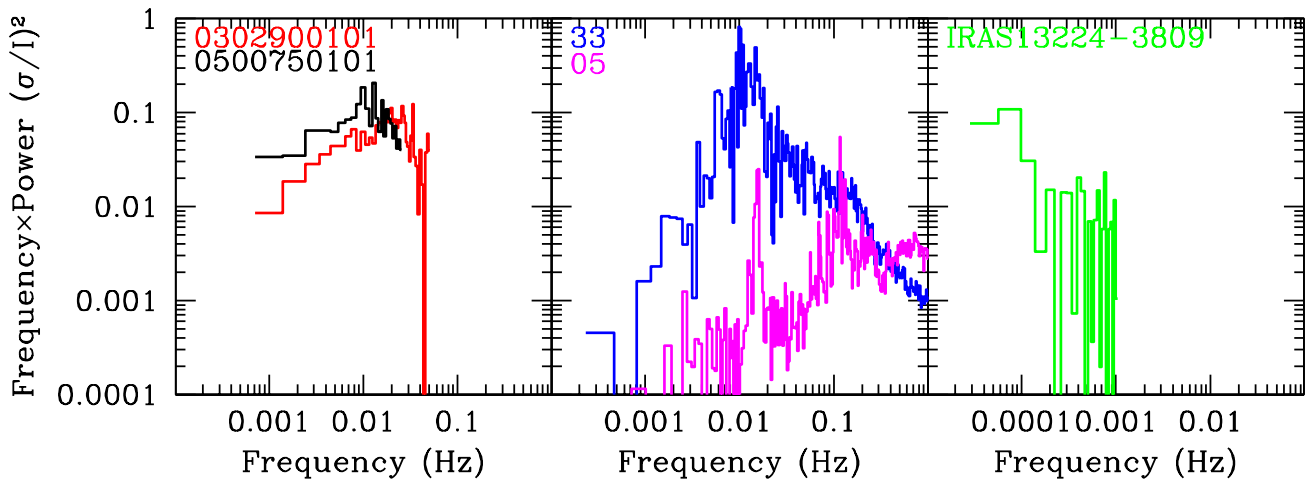}
\end{tabular}
\end{center}
\caption{Left-hand panel: Average (from 1000 s segments of the
lightcurve) PDS of both observations of NGC 5408 X-1 in the 1-8~keV
band. Centre panel: 2 PDS from \textit{RXTE} observations of the BHB
GRS 1915+105 in a highly luminous (probably super-Eddington) state
(OBSIDs: 20405-01-33-00 and 10408-01-05-00, termed 33 and 05, in blue and magenta respectively) created from the
entire length of the observation. Right-hand panel: the
\textit{XMM-Newton} EPIC-pn PDS of the NLS1 IRAS 13224-3809 created
from the entire length of the observation. All PDS are very slightly geometrically
re-binned (by a factor of 1.02) and have had their white noise
subtracted. The amount of variability (i.e. the integrated power under the PDS) seen in the tail of NGC 5408 X-1
is unusually high, and appears to only be matched by that seen in some
super-Eddington sources.}
\label{}
\end{figure*}

We can further resolve the variability by looking at long and short
timescales. We recalculate the rms using a binning of 50~s, to show
only the power on long timescales (red points). This shows a very
similar pattern of variability to that derived from the total
lightcurve, with much more variability at hard energies than at soft.
The blue points show the energy dependence of the rapid variability,
formed from the difference (in quadrature) between $\sigma/I(E)$ from
the 10~s binning and that from the 50~s binning. Thus this effectively
isolates variability between frequencies of 0.01-0.05~Hz, but again
the pattern of variability with energy is very similar. 

Both observations show a very similar pattern, but the rms variability
at 1-2~keV is even larger for Obs 2 than Obs 1. The black points in
Fig 6 show the ratio of the total rms spectrum in Obs 2 compared to
Obs 1, while the red and blue points show the ratio of rms for the
long and short timescale variability respectively. Apart from on short
timescales, these show significantly more variability above 1~keV in
Obs 2, as expected from the power spectral analysis (Fig 4).

Thus, variability on all timescales and in both observations shows a
strong increase as a function of energy. This is consistent with a
model in which there is a relatively stable component at soft energies
diluting the variability as might be expected by a thermal disc (see
Churazov et al. 2001) or optically thick photospheric component (G09,
Zdziarski, Misra \& Gierli{\'n}ski 2010). This can be modelled by
changing the normalisation of the hard energy spectral component by
the percentage rms and taking the ratio of this to the best-fit
model. The result on 10s binning is shown as the solid black line
in Fig 5, which closely resembles the shape of the rms. 

We note that the shape of this rms is inconsistent with models in
which the spectrum is dominated by a single component which changes
only in normalisation. This would give an rms which is constant with
energy. This rules out the reflection dominated model of
Caballero-Garcia \& Fabian (2010) unless the variability is strongly
fine-tuned. The rms shape requires that the unseen illuminating
power-law spectrum pivots around 0.3~keV (see the spectral and rms
analysis of a reflection dominated model for RE J1034+396 in Middleton
et al.  2009).

\begin{figure*}
\begin{center}
\begin{tabular}{l}
\leavevmode \epsfxsize=14cm \epsfbox{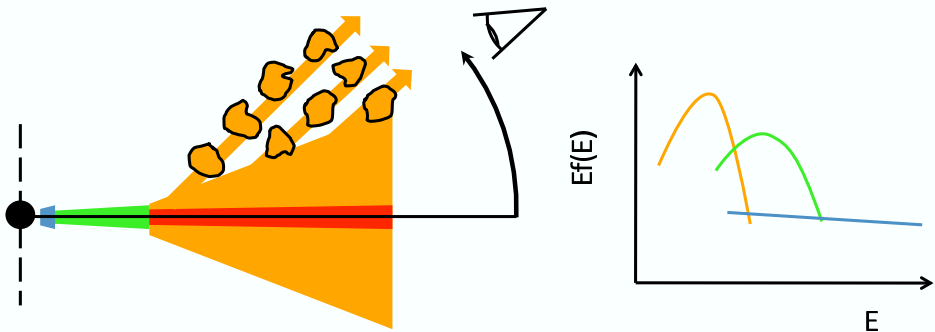}
\end{tabular}
\end{center}
\caption{Cartoon of a physical model for high mass accretion rate
sources. Here the quasi-thermal soft emission originates from a disc
(red) distorted by a photosphere (orange) which is the base of the
radiatively-driven wind. The green region indicates optically thick
Comptonisation which is stable and blue indicates the most energetic
region which is highly variable, appearing as optically thin
Comptonisation or a power-law to higher energies. This is outside of
the bandpass of ULXs as seen with {\it XMM-Newton} but is seen in GRS
1915+105 and certain NLS1s where the optically thick Comptonisation
can be seen to dilute the variability. This model suggests that, in
the ULXs where variability is suppressed, we are viewing them directly
along an unobscured line-of-sight to the central regions (hence they
are close to face-on), but those which are highly variable such as NGC
5408 X-1, the obscuration of the optically thick, Comptonised emission
region by a clumpy wind creates large scale variability.}
\label{}
\end{figure*}

\section{Discussion}

The spectral and rms analysis presented here both support a
two-component model, where the soft component can be modelled by low
temperature thermal emission that is mostly constant on timescales of
up to a day, while the hard component, best modelled by a cool,
optically thick corona, is steep and varies quite dramatically on
timescales down to at least a hundred seconds. Whilst this
superficially resembles the properties of the 'steep power-law' state
(McClintock \& Remillard 2006) seen at fairly high luminosities, both
the spectra and variability are inconsistent with such an
interpretation. The low temperature of the thermal component means
that if this were from an accretion disc then the black hole mass is
of order $10^4 M_\odot$. A similarly large mass is derived if the QPOs
are identified with the BHB type C LFQPO. However, the ULX luminosity
of $\sim 10^{40}$~ergs s$^{-1}$ is then only $L/L_{Edd}\sim
0.01$. BHBs have hard tails ($\Gamma\sim 1.5-1.8$) at such low mass
accretion rates, rather than the soft tail\footnotemark seen in NGC
5408 X-1. Similarly, the BHB power spectra at these low mass accretion
rates show a low frequency break tightly correlated with the QPO
frequency. These breaks are not present at the predicted frequency in
NGC 5408 X-1.

\footnotetext{The negative slope of the data at E$>$1.5 keV in both
panels of Fig. 2 demonstrates that $\Gamma>$2, assuming a power-law
fit, given the panels are plotted in Ef(E) space.}

Soft, variable tails are only seen in BHBs in the 'steep power-law'
state, at $L/L_{Edd}\sim 0.3-1$ (Remillard \& McClintock 2006), giving
a mass of $80-230 M_\odot$ if NGC 5408 X-1 is in an analogous state.
However, the large fractional variability of 0.3-0.4 is only seen
above 10~keV in BHBs in this state, where the tail dominates over the
constant disc component (see the energy-dependent rms spectra in
Gierli{\'n}ski \& Zdziarski 2005). Scaling this energy for the
difference in mass from $80-230 M_\odot$ to $10M_\odot$ gives a
prediction that the tail should start to dominate above 4.5-6~keV, at
significantly higher energies than seen in NGC 5408 X-1, where the
variable tail dominates above 2~keV.

Thus, NGC 5408 X-1 does not look convincingly similar to any of the
sub-Eddington BHB states. Instead, the large
implied emission radius of the soft thermal component can also be
incorporated into the super-Eddington flow models by identifying this
not with the disc itself but with much more extended emission from the
photosphere of a wind (Shakura \& Sunyaev 1973; Begelman et al. 2006;
Poutanen et al.  2007). Similarly, the 
extreme variability associated with the tail can be matched by
occasional states of the pathologically variable BHB, GRS 1915+105,
where it is probably associated with instabilities in a
super-Eddington flow (Belloni et al. 1997).  Fig 7 (left hand panel)
shows the power spectrum of the tail in NGC 5408 X-1 (1-8~keV)
compared to that of {\it RXTE} data from $\kappa$ and $\mu$ variability
states of GRS 1915+105 (centre panel, OBSIDs 20405-01-33-00: Belloni et al. 2000).  These are calculated using
standard 1 data (0.125 s resolution) which has no energy information,
but the source spectrum in the {\it RXTE} bandpass means that the counts are
concentrated in the 3-10~keV energy range. As well as showing
similarly extreme amounts of broad band noise, these also show small
milli-Hertz QPO features at similarly low frequencies as seen in NGC
5408 X-1 (0.01~Hz for 33).  These ultra-low frequency
QPOs can dominate the power spectrum of GRS 1915+105 in other states
where there is much less broad band variability (OBSID: 10408-01-05-00
shown in magenta in the same panel), but this is unlike NGC 5408 X-1,
where the QPOs are weak and the broadband variability is
strong. Nonetheless, it is clear that stellar remnant black holes can
produce ultra-LFQPOs as well as the standard type C LFQPOs.  

However, while the variability properties of NGC 5408 X-1 and some
states of GRS 1915+105 are very similar, the energy spectra are subtly
different. The $\kappa$ and $\mu$ state observations of GRS 1915+105
have spectra which can be well modelled by a disc plus low
temperature, optically thick Comptonisation (Zdziarski et al. 2001,
2005; Done,  Wardzi{\'n}ski \& Gierli{\'n}ski 2004; Ueda et al. 2009, 2010), but
{\em unlike} NGC 5408 X-1, the Comptonisation component dominates the
bolometric luminosity, rather than the thermal component (see Fig. 2,
compared to Middleton \& Done 2010).

Thus, even the most super-Eddington BHB is not really a good match to
the spectrum of NGC 5408 X-1. The only known systems which show a
similar combination of steep spectral tail and extreme variability are
a small subset of Narrow Line Seyfert 1 galaxies (Leighly 1999;
O'Neill et al.  2005) \footnotemark \footnotetext{However, we note that a second ULX with similar properties was reported while this paper was being reviewed (Rao, Feng \& Kaaret 2010)}. These are likely to be super-Eddington sources,
but there are large uncertainties in deriving mass and mass accretion
rate for these sources (Grupe \& Mathur 2004; Marconi et al. 2008).
Fig 7 (right hand panel) shows the power
spectrum from {\it XMM-Newton} data (OBSID 0110890101) of the most extremely
variable NLS1, IRAS 13224-3809
(Boller et al. 1997; Gallo et al. 2004;
Ponti et al. 2010).  Since any soft thermal component in the AGN should be
at much lower energies, we use the full bandpass from
0.3-10~keV. Clearly the peak in $\nu P(\nu)$ reaches similar values,
though the frequencies are very different.

Since NGC 5408 X-1 does not match with any of the sub-Eddington BHB
states, nor to any of the multiple states of the mildly
super-Eddington BHB GRS 1915+105, it seems most likely that it
represents a strongly super-Eddington source. Such flows power a
strong wind, where the wind photosphere gives the large radius,
luminous, soft thermal component at low energies (Shakura \& Sunyaev
1973; Begelman et al. 2006; Poutanen et al. 2007). This extended
emission will be constant on short timescales. We also see a much more
variable tail, so the optically thick wind cannot cover all lines of
sight to the inner disc. A reasonable geometry for this is an
equatorial disc wind (Abolmasov, Karpov \& Kotani 2009; Ohsuga et al. 2007, 2009).

The low temperature, optically thick Compton emission could be
produced by a corona over the inner disc, a more extreme version of
the marginally optically thick corona seen in the steep power-law
state (e.g. Done \& Kubota 2006; G09).  However, it could also be the
spectrum of the inner disc itself as at such high mass accretion rate,
the high disc temperature means that there is very little true
continuum opacity (Beloborodov 1998, Done \& Davis 2008). This means that the energy
does not thermalise, and cannot even be well approximated by a colour
temperature corrected blackbody (see e.g. Watarai et al. 2001). We
favour this latter origin as the probably super-Eddington AGN RE
J1034+396 shows clear evidence for a {\em separate}, variable corona,
as well as a more constant low temperature Comptonised component (the
energy shift from the much larger black hole mass means that any wind
photosphere emission is at too low an energy to be observed). This
predicts there should also be an additional weak tail in the ULX,
which is so far unobserved.

While the small size scale of the inner disc means that it could vary
intrinsically, we note that other ULXs with similar spectra do not show
similarly extreme variability (Heil et al. 2009, spectra from
G09). Instead, a solution which can explain the range of variability
behaviour in these wind-dominated ULXs could be if the extreme
variability is extrinsic, from clumps in the wind. Numerical
simulations show that this wind is turbulent (Ohsuga 2007; Ohsuga et
al. 2009), giving strong variability of the Comptonisation component
from the inner disc on sightlines which intercept the major wind
streamlines, but the soft thermal emission will remain mostly constant
due to its much larger size scale. 

We show a simple cartoon of this model in Fig 8 (see also Leighly
2004). At extreme super-Eddington mass accretion rates, a large
fraction of the accretion energy is ejected in the wind, making this
energetically dominant (Poutanen et al. 2007).  The photosphere of the
wind (orange) is the source of soft thermal emission, while the inner
disc (green) is the source of the low temperature, optically thick
Comptonised emission.  We propose that this has only low level
intrinsic variability in strongly super-Eddington sources, perhaps
because the disc structure is stabilised by advection (Abramowicz et
al. 1988). Thus these sources have very low variability when viewed
along a line of sight which does not intercept the wind, while there
is strong variability from obscuration of the Comptonised emission when
viewed through the turbulent wind. The intrinsically variable, but
rather weak corona (blue) gives rise to a weak high energy tail, so
far not observed in ULXs.  

Scaling this model up to NLS1s could perhaps similarly explain the
extreme variability from sources such as IRAS 13324-3809 and
1H0707-495 from absorption in strong winds (Gallo 2004, see also
Miller et al. 2010 for how this can also match the spectrum), the
existence of which are supported by UV emission lines (Leighly
2004). The much lower variability seen from the low temperature
Comptonised component in RE J1034+396 is then associated with a more
face-on viewing angle.

\section{Conclusion}

Where timing characteristics scale with mass, they can be used to
determine the unknown mass of the black hole in ULX. However, the wide
variety of properties seen in BHB means that this only works if the
correct spectral state is identified for scaling. The key issue in ULXs
is whether they are analogous to any of the classic, sub-Eddington,
BHB states.  If so, the large luminosities of these systems implies an
IMBH. Instead, if ULXs are super-Eddington, then there are no well
defined templates from BHB except for the unique source GRS 1915+105.

The suggested identification of the QPO in NGC 5408 X-1 is with the type C
LFQPO seen in BHBs. However, BHBs show a tight correlation between
this QPO frequency and that of the low frequency break in the broad
band power spectrum (Wijnands \& van der Klis 1999). We use extensive
numerical simulations to constrain the position of this break in the
two observations of NGC 5408 X-1, and find that they are inconsistent with
the predicted frequencies from a type C LFQPO.

The energy spectra of both observations of NGC 5408 X-1 are best described
by a soft thermal component (disc or photosphere) together with a low
temperature, optically thick, thermal Compton component, as proposed
for other high quality ULX spectra (G09). The rms spectra show that
the variability increases strongly with energy, consistent with a
constant soft component which dilutes a separate, strongly variable
component at higher energies. The amount of variability in the high
energy component is extreme, and we speculate that this is extrinsic,
from stochastic obscuration by the turbulent wind.  The only objects
with comparable variability and spectra are GRS 1915+105 and some
extreme NLS1s, both of which are most probably super-Eddington
accretors. This also suggests an alternative identification for the
QPO in NGC 5408 X-1 as being more closely related to the ultra low
frequency QPO seen occasionally in GRS 1915+105. Both spectra and
timing are then consistent with identifying NGC 5408 X-1 as a
super-Eddington source. This places a natural upper limit on the
central BH mass of $\sim$ 100M$_{\odot}$, consistent with a BH formed
from stellar collapse in a low metallicity environment
(e.g. Belczynski et al. 2010).

\section{acknowledgements}

We thank the referee, Juri Poutanen, for useful comments and insights. MM thanks STFC
for their support in the form of a post-doctoral position funded by a standard grant. This work is based on observations obtained with {\it XMM-Newton}, an ESA science mission with instruments and contributions directly funded by ESA Member States and NASA.

\label{lastpage}

\end{document}